\documentclass[12pt]{iopart}
\usepackage{graphicx}
\begin{document}
	\title[Non-perturbative effects in corrections to quantum master equations]{Non-perturbative effects in corrections to quantum master equations arising in Bogolubov-van Hove limit}
	\author{Alexander E Teretenkov}
	\address{Department of Mathematical Methods for Quantum Technologies,\\ Steklov Mathematical Institute of Russian Academy of Sciences,\\ ul.\,Gubkina 8, Moscow 119991, Russia.}
	\ead{\mailto{taemsu@mail.ru}}

	\begin{abstract}
		We study the perturbative corrections to the { Gorini-Kossakowski-Sudarshan-Lindblad equation which arises in the  weak coupling limit}.  The spin-boson model in the rotating wave approximation at zero temperature is considered. We show that the perturbative part of the density matrix satisfies the time-independent Gorini-Kossakowski-Sudarshan-Lindblad equation for arbitrary order of the perturbation theory (if all the moments of the reservoir correlation function are finite).  But to reproduce the right asymptotic precision at long times, one should use { an initial condition different} from the one for exact dynamics. Moreover, we show that the initial condition for this master equation even fails to be a density matrix under certain resonance conditions.
	\end{abstract}


		\section{Introduction}
	
	Due to modern interest  in the non-Markovian phenomena in the open quantum systems (see, e.g., \cite{Breuer09, Rivas14, Chruscinski2019, Li2020}) there is a widening range of approaches to their description (see, e.g., \cite{Filippov2018, Luchnikov19, Luchnikov2020, Trushechkin19, Hsieh2019, Whalen2019, Budini2019} for a few recent ones). But here we consider possibly the most direct one, namely,  the derivation of perturbative corrections to Markovian master equations \cite{Jang2002, Trushechkin19a}. A widespread approach to perturbative derivation of quantum master equations in the weak coupling limit is based on expansion of the Nakajima-Zwanzig equation \cite{Nakajima1958, Zwanzig1960} with a coupling constant as a small parameter for the fixed time (see \cite[section~9.1-9.2]{Breuer02}, \cite[section~3.8]{May2008}, and \cite[section~1.3]{Carmichael2013}).   However, the most mathematically strict derivations \cite{Davies1974}, \cite[section~10.5]{davies1976quantum}, \cite{Accardi2002} are based on the Bogolubov-van Hove scaling \cite{Bogoliubov1946, Bogoliubov1955}. { This means that one introduces a small parameter $ \lambda $ before the interaction Hamiltonian and also rescales the time as  $ t \rightarrow \lambda^{-2} t  $. So we have a very specific combination of weak coupling and long-time expansion if we consider perturbation theory in this small parameter. The limit $ \lambda \rightarrow 0 $ is  called the Bogolubov-van Hove limit.}  The  equations of  the  Gorini-Kossakowski-Sudarshan-Lindblad (GKSL)  form  {arise only after this limit} \cite{davies1976quantum, Accardi2002}, which leads to the completely positive (CP) trace preserving dynamics of the density matrix \cite{gorini1976completely, lindblad1976generators}. { Moreover, the  GKSL equations which occur in this case are deeply studied \cite{Accardi2016, Fagnola2019, Cervantes2020}. In these works they are called weak coupling limit type (WCLT) GKSL equations. So, we will also use this term.} In particular, the initial density matrix remains a density matrix at any time. This is not the case for integro-differential equations in the Born approximation or for the Redfield equation \cite{Redfield1957} which arises in the approach without { the time} scaling, when the initial density matrix can become non-positive during the dynamics \cite{Dumcke1979, Suarez1992, Zhao2002, Benatti2003, Whitney2008, Hartmann2020}.  In the Bogolubov-van Hove limit besides reduced density matrix dynamics one could  also obtain stochastic unitary dynamics of the system and the reservoir \cite{Accardi2002}{. In particular, one obtains} Markovian dynamics of the system multitime correlations. It differs from the possible situation, when the reduced density matrix satisfies the GKSL master equation, but the dynamics of multitime correlation functions is non-Markovian \cite{Gullo2014, Strasberg2018, Li2018, Milz2019}. In such a sense the Bogolubov-van Hove limit leads to { the} true Markovian behavior in the zeroth order of perturbation theory.  
	
	From the physical point of view the scaling $ t \rightarrow \lambda^{-2} t  $ of the  Bogolubov-van Hove limit allows one to separate the time scale on which the Markovian  behavior occurs (which is defined by finite $ t $ after  { the} scaling) from the time scale of order of reservoir correlation time (which  is defined by $ t = O(\lambda^2) $ after scaling and tends to zero in the limit $ \lambda \rightarrow 0 $). We refer to the behavior on the former time scale as long-time behavior and on the latter time scale as short-time one throughout the article.
	
	The non-Markovian behavior at short times of order of the bath correlation time is not surprising.  (Sometimes the scale of a short initial non-Markovian period of time is  called quantum Zeno time \cite{Petrosky2002}.) But it is interesting to obtain perturbative corrections to the WCLT GKSL equation to understand long-time { behavior of the system}. A general approach to that  based on multipole noises was developed in \cite{Pechen2002, Pechen2004}{. These works were in turn based on the results of \cite{Arefeva2000}. But \cite{Pechen2002, Pechen2004} were focused} on deriving stochastic equations for both the system and the reservoir. However, our work is focused on the dynamics of the reduced density matrix of a very special paradigmatic system, namely, { on the spin-boson model} at zero temperature in the rotating wave approximation (RWA). This system is well-studied \cite{Imamoglu94, Garraway96, Garraway97, Garraway97a, Dalton01, Garraway06, Teretenkov19m} and widely used as a basic example \cite{Breuer1999, Haikka2011, Gullo2014, Lima2020}. So we choose to study the perturbation theory with the Bogolubov-van Hove scaling { using this model}. We also do not discuss the validity of the RWA which was called into question in the literature \cite{Fleming10, Benatti10, Ma2012, Tang13, Trubilko2020}. 
	
	{
		Putting aside some technical details, which are presented in the main text, let us stress the main results of this paper
		\begin{enumerate}
			\item \label{it:reduced}
			After the reservoir correlation time the dynamics of the reduced density matrix could be described by the GKSL equation with constant coefficients with high precision. 
			\item \label{it:corr}
			After the  correlation time the system  two-time correlation function satisfies the Markovian formula (i.e. the quantum regression formula) with minor correction by the constant multiplier with the same precision.  
			\item The asymptotic precision of such approximation is fully characterized by the number of the finite moments of the bath correlation function. In the case, when all moments of the bath correlation function are finite, our approximation is valid in all orders of perturbation theory. Only the terms which turn to zero faster than any power of $ \lambda $, which we call non-perturbative ones, are omitted in this case by our approximation.
			\item The initial condition at $ t=0 $ for our GKSL equation should be different from the one for exact dynamics. Moreover, in general this initial condition should be not physical, i.e. not a density matrix under the resonance conditions discussed in section~\ref{sec:cond}.
		\end{enumerate}
		
		Let us also discuss what seems for us to be novel in our results from several viewpoints of research.
		
		Firstly, we regard  \eref{it:reduced}--\eref{it:corr} as a definition of long-time Markovianity. This approach to Markovianity is new, in the sense that it is not presented in the modern hierarchy of definitions of Markovianity \cite{Li2018, Milz2019}. In particular, for the explicit example considered in section \ref{sec:explicit} the   Breuer-Laine-Piilo (BLP) \cite{BreuerLaine09} and Rivas-Huelga-Plenio (RHP) \cite{RivasHuelga14} measures of non-Markovianity classify its dynamics as Markovian inside the convergence radius in $ \lambda $. The measure based on system correlation functions \cite{Gullo2014} regards  it as non-Markovian under the same conditions. From our point of view we have something in between: it is non-Markovian at all the times, but only long-time Markovian. Let us also note that even for long times the trace of non-Markovianity lasts in the possible range of the initial conditions, so it is impossible to say that all the non-Markovian effects are forgotten after the correlation time. In this sense one could say that the long-time Markovianity is also between Markovian and non-Markovian behavior.
		
		Secondly, our GKSL equation is time-independent and, hence, simpler in this case than the standard time-convolutionless master equation \cite[section~9.2]{Breuer02} derived by cumulant expansion techniques \cite{Kubo1963, vanKampen1979}, but it provides the same asymptotic precision for long-time dynamics. { However,} the price for this simplification of equations consists in  difference between the initial condition for the exact and asymptotic equation.
		
		This difference is a well-known phenomenon of singular perturbation theory (\cite[section~7.2]{Bender1978} and \cite[section~1.3]{Lagerstrom1988}) and is sometimes called the initial layer phenomenon. The discussion of this phenomenon for some other physical models could be found in  \cite{VanKampen1985}. The singularity of the perturbation theory here coincides with \cite{Meng19}, where multiple-scale singular perturbation theory was used for open quantum systems.
		
		So, thirdly, from the point of view of initial layer and slippage phenomena we additionally show that for this specific model one should redefine the initial condition in a non-physical way in some cases to obtain the right asymptotic precision.	This fact is actually { extremely} contraintuitive as we have shown that we have the GKSL equation for the density matrix, so there is nothing in the equation itself, why one should use non-physical initial condition for this equation. The situation here is inverse of the one in the case of the slippage for the Redfield master equation \cite{Suarez92, Gaspard99, Whitney08, Rivas17}. There one should restrict the initial conditions to obtain positive evolution. So for uncorrected initial conditions the Redfield master equation gives non-physical evolution after some times, { therefore} one could understand the necessity of this correction by the equation itself. For our case if one takes the uncorrected initial condition, it is physical at any time, but it gives wrong asymptotic precision (see, in particular, figure~\ref{fig_0} for the explicit example).
		
		Fourthly, in the { zeroth} order of the perturbation theory with the Bogolubov-van Hove scaling there is no initial layer phenomenon and one obtains the GKSL equation which is valid uniformly in time, but for further corrections we show that this is not the case. It seems to be the main reason for { a} very small amount of works about the perturbative master equation in case of { the} Bogolubov-van Hove scaling. Moreover, the known multipole expansion \cite{Pechen2002, Pechen2004}  considers the limit of dynamics both for the system and reservoir, so all the derivatives of the delta-function contribute to their approximation of the bath correlation function. But we show that from the point of view of the  system dynamics the reservoir correlation functions could be approximated just by the single delta function with the same asymptotic precision. 
	}
	
	\section{Asymptotic behavior of zero-temperature RWA spin-boson}
	\label{sec:assymptBehav}
	
	We consider the model of spin-boson in the rotating wave approximation at zero temperature due to the fact that its reduced density matrix could be obtained exactly in terms of a scalar integro-differential equation.  Namely, we consider the two level system with { the} ladder operators $ \sigma_+ \equiv |1\rangle \langle 0|, \sigma_- \equiv |0\rangle \langle 1| $ and a bosonic reservoir with creation and annihilation operators $ b_k, b_k^{\dagger} $ satisfying canonical commutation relations $ [b_k, b_{k'}^{\dagger}] = \delta(k - k') $, $ [b_k, b_{k'}] = 0  $. Then the Hamiltonian of the spin-boson in RWA has the form 
	\begin{eqnarray}\label{eq:SBHam}
		H = \int \omega(k)  b_k^{\dagger} b_k d k + \Omega \sigma_+ \sigma_- +  \int \left(  g^*(k)  \sigma_- b_k^{\dagger}+   g(k)   \sigma_+  b_k \right) d k.
	\end{eqnarray}
	The evolution of the density matrix is described by the Liouville--von Neumann equation
	\begin{equation}
		\frac{d}{dt} 	\rho_{SB}(t) = - i[H, \rho_{SB}(t) ].
	\end{equation}
	We consider the zero-temperature initial condition $ \rho_{SB}(0) =  \rho(0) \otimes | \rm{vac} \rangle \langle \rm{vac}| $, where $ | \rm{vac} \rangle $ is the vacuum vector: $ b_k  |\rm{vac}\rangle =0 $. We are interested in the dynamics of the reduced density matrix in the { interaction} picture which takes the form $ \rho(t) \equiv e^{i \Omega \sigma_+ \sigma_- t} \Tr_{B} \rho_{SB}(t) e^{-i \Omega \sigma_+ \sigma_- t}  $ in this case (taking into account that $ | \rm{vac} \rangle \langle \rm{vac}|  $ is invariant under free evolution).
	
	Then (see \cite[p.~463]{Breuer02} for pure states and \cite{Teretenkov19} for arbitrary density matrices) the reduced density matrix dynamics could be represented in the form
	\begin{eqnarray}\label{eq:demMatEvol}
		\rho(t) = \left( \begin{array}{cc}
			|x(t)|^2 \rho_{11}(0) & x(t) \rho_{10}(0)\\
			x^*(t) \rho_{01}(0) & \rho_{00}(0) + (1-|x(t)|^2 )\rho_{11}(0) 
		\end{array} \right),
	\end{eqnarray}
	where $  x(t) $ is a (unique) solution of the integro-differential equation
	\begin{equation}\label{eq:intDiffIntPic}
		\frac{d}{dt} x(t) =  -\int_0^t d\tau G(t - \tau)x(\tau), \qquad x(0) = 1,
	\end{equation}
	where
	\begin{equation}
		G(t) \equiv \int d k e^{-i (\omega(k) - \Omega)  t} |g(k)|^2.
	\end{equation}
	If we now introduce a small parameter scaling into the coupling constant $ g(k) \rightarrow \lambda g(k)$, then \eref{eq:intDiffIntPic} takes the form
	\begin{eqnarray}
		\frac{d}{dt} x_{\lambda}(t) =  -  \lambda^2 \int_0^t d\tau G(t - \tau)x_{\lambda}(\tau).
	\end{eqnarray}
	We are going to work with $ x_{\lambda}(t) $ after the scaling $ t \rightarrow t/\lambda^2 $, i.e. with $ x(t; \lambda) = x_{\lambda}(\lambda^{-2} t) $. Rewriting the previous equation in terms of $ x(t; \lambda) $ we have
	\begin{eqnarray}
		\lambda^2\frac{d}{dt} x(t; \lambda) = - \lambda^2 \int_0^{ t/\lambda^2} G\left(\frac{t}{\lambda^2} -\tau\right) x(\lambda^2 \tau ; \lambda) d\tau.
	\end{eqnarray}
	After change of the variable $ s = \lambda^2 \tau $ we obtain
	\begin{equation}\label{eq:mainIntDiff}
		\frac{d}{dt} x(t; \lambda) = - \int_0^t \frac{1}{\lambda^2} G\left(\frac{t-s}{\lambda^2}\right) x(s ; \lambda) ds,
	\end{equation}
	where $  x(0; \lambda)=1 $. Similarly, the scaled dynamics of the density matrix $ \rho(t; \lambda) $ is defined by \eref{eq:demMatEvol}, where $ x(t) $  is replaced by $ x(t; \lambda) $. Let us note that asymptotic analysis of \eref{eq:mainIntDiff} without the Bogolubov-van Hove scaling is discussed in \cite{Bologna10}.
	
	The limit $ \lambda \rightarrow 0 $ corresponds to the secular Redfield equation (the Redfield and the secular Redfield equations coincide in this case due to RWA in the initial Hamiltonian \cite{Teretenkov19}). So let us focus on the first asymptotic correction of the solution of \eref{eq:mainIntDiff} (see \ref{app:Exp} for derivation of formulae for perturbative terms of arbitrary order):
	\begin{equation}\label{eq:xSecOrder}
		x(t; \lambda) = e^{-\tilde{G}_0 t} \left(1 - \tilde{G}_1 (1 - \tilde{G}_0 t) \lambda^2 + O(\lambda^4)\right),
	\end{equation}
	where 
	\begin{equation}
		\tilde{G}_0 =\int_0^{\infty} G (t) dt, \qquad \tilde{G}_1 = -\int_0^{\infty} t  G (t) dt.
	\end{equation}
	Taking into account \eref{eq:demMatEvol}, we obtain 
	\begin{equation}\label{eq:secOrderDenMat}
		\eqalign{
			\rho_{11}(t;\lambda) &\simeq  (1 -2 \Re(\tilde{G}_1 (1 - \tilde{G}_0 t)) \lambda^2) e^{- 2(\Re \tilde{G}_0) t} \rho_{11}(0),  \\
			\rho_{10}(t;\lambda) &\simeq  (1 - \tilde{G}_1 (1 - \tilde{G}_0 t) \lambda^2) e^{-\tilde{G}_0 t} \rho_{10}(0), 
		}
	\end{equation}	
	where terms of order $  O(\lambda^4)  $ are neglected.
	
	As we have discussed in the introduction, we call the asymptotic expansion of $ \rho(t) $ for $ t>0 $ in powers of $ \lambda $ a perturbative part of $ \rho(t) $, so we denote it $  \rho(t)|_{\rm pert} $. We will also use the same notation for other functions of $ \lambda $ and $ t $. Generally, $ \rho(t)|_{\rm pert} \neq \rho(t)$. To show it, let us take the limit $ t \rightarrow +0 $ for $ \rho(t)|_{\rm pert} $. By \eref{eq:secOrderDenMat} we obtain
	\begin{equation}\label{eq:initCondPert}
		\eqalign{
			\rho_{11}(+0;\lambda)|_{\rm pert} &=  (1 -2 \Re(\tilde{G}_1 ) \lambda^2)  \rho_{11}(0) + O(\lambda^4),\\
			\rho_{10}(+0;\lambda)|_{\rm pert} &=   (1 - \tilde{G}_1  \lambda^2) \rho_{10}(0)  + O(\lambda^4). }
	\end{equation}	
	Hence,  the initial conditions for the perturbative part $ \rho(+0;\lambda)|_{\rm pert} $ of the density matrix generally do not coincide with the initial condition $ \rho(0) $ for the whole density matrix. Moreover, to reproduce true asymptotic expansion, these initial conditions have to be non-physical if $ \Re \tilde{G}_1 < 0 $. 
	
	If one wants to understand this difference between the initial condition for the whole density matrix and its perturbative part by a simple explicit example, one should refer to section~\ref{sec:explicit}.
	
	\section{Asymptotic GKSL equation and long-time Markovianity}
	\label{sec:assympInfDiv}
	
	For $ x(t) \neq 0 $  density matrix \eref{eq:demMatEvol} satisfies a time-local master equation 
	\begin{equation}\label{eq:masterEq}
		\frac{d}{dt} \rho(t) = \mathcal{L}_t(\rho(t))
	\end{equation}
	with a generally time-dependent GKSL-like generator \cite[p.~463]{Breuer02}
	\begin{equation}
		\mathcal{L}_t(\rho) = - i[ \Delta \Omega(t) \sigma_+ \sigma_-, \rho]+\Gamma(t) \left(\sigma_- \rho(t) \sigma_+ - \frac12 \sigma_+\sigma_- \rho - \frac12 \rho \sigma_+\sigma_- \right), \label{eq:likeGKSL}
	\end{equation}
	where
	\begin{equation}\label{eq:likeGKSLParam}
		\Delta \Omega(t) = - \Im \frac{\dot{x}(t)}{x(t)}, \qquad \Gamma(t) = -2 \Re \frac{\dot{x}(t)}{x(t)}.
	\end{equation}
	It is an actual time-dependent GKSL generator in the case { when} $ \Gamma(t) \geq 0 $ for all $ t \geq 0 $. In such a case the dynamics is CP-divisible \cite{Hall2014}. Asymptotic expansion \eref{eq:xSecOrder} leads to 
	\begin{equation}
		\Gamma(t)= 2 \Re \tilde{G}_0  (1  -  \tilde{G}_1 \lambda^2) + O(\lambda^4).
	\end{equation}
	Hence, $ \Gamma(t) $ asymptotically does not depend on time. Actually, it is possible to prove under general conditions (see \ref{app:pertPartDiffEq}) that 
	\begin{equation}\label{eq:xPert}
		x(t;\lambda)|_{\rm pert} = r(\lambda) e^{\tilde{p}(\lambda) t},
	\end{equation}
	where $ r(\lambda) $ and $ \tilde{p}(\lambda) $ are time-independent functions of $ \lambda $. Then both perturbative parts
	\begin{equation}
		\Delta \Omega(t)|_{\rm pert} = - \Im \tilde{p}(\lambda),\qquad \Gamma(t)|_{\rm pert} = -2 \Re \tilde{p}(\lambda)
	\end{equation}
	are time-independent. Hence, to describe asymptotic dynamics of the reduced density matrix, one could use  a master equation with the time-independent  GKSL generator, but with the generally non-physical initial condition:
	\begin{equation}
		\rho(+0)|_{\rm pert}  = 
		\left( \begin{array}{cc}
			|r(\lambda)|^2 \rho_{11}(0) & r(\lambda) \rho_{10}(0)\\
			r^*(\lambda) \rho_{01}(0) & \rho_{00}(0) + (1-|r(\lambda)|^2 )\rho_{11}(0) 
		\end{array} \right).
	\end{equation}
	A similar phenomenon arises in \cite{Basharov2019}, where the corrections based on algebraic perturbation theory also led to the equation which still has the GKSL form. Also, our time-independent GKSL equation is an analog of kinetic equations for oscillator considered in \cite[section~IV]{Petrosky2002}, where  the Bogolubov-van Hove scaling of time was also used.
	
	Besides this initial condition the dynamics has the GKSL form and is Markovian in this sense. But there is a discussion in literature about the definition of quantum Markovianity which leads to the hierarchy of generally nonequivalent definitions of Markovianity  \cite{Li2018, Milz2019}. In particular, there is a point of view that both in the classical \cite{VanKampen1992} and the quantum case \cite{Gullo2014, Strasberg2018, Li2018, Milz2019} Markovianity does not affect only the master equation, but also affects multitime correlation functions. Namely, Markovian dynamics assumes that two-time correlation functions \cite{Gullo2014} have a very special form defined by the quantum regression formula (see, e.g., \cite[section~5.2]{Gardiner2004} and  \cite[subsection~3.2.4]{Breuer02})
	\begin{equation}
		\langle \sigma_-(t_2) \sigma_+(t_1)\rangle_{\rm M}= \Tr (\sigma_- \Phi_{t_1}^{t_2}(\sigma_+\Phi_{0}^{t_1}(|0\rangle\langle 0|))) = \frac{x(t_2)}{x(t_1)},
	\end{equation}
	where $ \Phi_{t_1}^{t_2} $ is defined by evolution of the density matrix as $ \rho(t_2) = \Phi_{t_1}^{t_2} \rho(t_1), t_2 \geq t_1 $. 
	In our case from \eref{eq:intDiffIntPic} we obtain explicitly
	\begin{equation}
		\Phi_{t_1}^{t_2}(\rho)=
		\left(\begin{array}{cc}
			\left|\frac{x(t_2)} {x(t_1)}\right|^2 \rho_{11} & \frac{x(t_2)}{x(t_1)} \rho_{10}\\
			\frac{x^*(t_2)}{x^*(t_1)} \rho_{01} & \rho_{00} +  \left(1 - \left|\frac{x(t_2)} {x(t_1)}\right|^2\right)\rho_{11} 
		\end{array}\right),
	\end{equation}
	and, hence,
	\begin{equation}\label{eq:MarkCorrFun}
		\langle \sigma_-(t_2) \sigma_+(t_1)\rangle_{\rm M}=  \frac{x(t_2)}{x(t_1)}.
	\end{equation}
	On the other hand, this correlation function can be calculated exactly (see \ref{app:corrFun})
	\begin{equation}\label{eq:exactCorrFun}
		\langle \sigma_-(t_2) \sigma_+(t_1)\rangle = x(t_2-t_1).
	\end{equation}
	Taking into account \eref{eq:xPert} we obtain
	\begin{equation}\label{eq:correlFunct}
		\langle \sigma_-(t_2) \sigma_+(t_1)\rangle_{\rm M} = r \langle \sigma_-(t_2) \sigma_+(t_1)\rangle, \qquad t_2 > t_1.
	\end{equation}
	So these correlation functions do not  coincide, but differ by time-independent factor $ r $. This factor occurs as a jump at initial time
	\begin{equation}
		\lim\limits_{\varepsilon \rightarrow +0}\frac{\langle \sigma_-(t + \varepsilon) \sigma_+(t)\rangle}{\langle \sigma_-(t) \sigma_+(t)\rangle} = r.
	\end{equation}
	If one renormalized this correlation function for long times as
	\begin{equation}
		\langle \sigma_-(t_2) \sigma_+(t_1)\rangle_{r} \equiv r \langle \sigma_-(t_2) \sigma_+(t_1)\rangle,
	\end{equation}
	then it would coincide with the Markovian one. In this sense the dynamics of this correlation function is also Markovian besides the initial jump. So it is possible to refer to such a behavior as long-time Markovian. But let us stress that this long-time Markovianity in terms of the renormalized correlation function
	\begin{equation*}
		\langle \sigma_-(t_2) \sigma_+(t_1)\rangle_{\rm M} = r \langle \sigma_-(t_2) \sigma_+(t_1)\rangle
	\end{equation*}
	is, strictly speaking, a different definition of Markovianity in terms of the correlation function than the usual one \cite{Gullo2014}, \cite[section~3.4]{Li2018} and they coincide only in the zeroth order of perturbation theory. (For the discussion of the zeroth order of perturbation theory in the far more general case see \cite{Dumcke1983}.) 
	
	To be valid to arbitrary order of perturbation theory, \eref{eq:xPert} needs the existence of all  the moments of $ G(t) $, otherwise it is valid accurate within a certain finite order  (see discussion at the end of \ref{app:Exp}). { Non-existence of the moments occurs, when $ G(t) $ decays slower than some (possibly fractional) power.} So it is natural that this interaction limits the possibility of the Markovian description  to the finite asymptotic order. To emphasize that { the} long-time Markovian behavior in all the orders of perturbation theory is non-universal,  in \ref{app:nonUniversality} we consider an example of a spectral density which leads to long-time Markovianity in the zeroth order of perturbation theory but fails to do it in the next order. We also show there that for arbitrary spectral density, which leads to { the} long-time Markovian behavior in the second order of perturbation theory, it is possible to find infinite number of spectral densities which lead to { the} long-time Markovian behavior only in the zeroth order of perturbation theory. So for almost all spectral densities such that the system dynamics is long-time Markovian in the zeroth order of perturbation theory it is not long-time Markovian in the next order of perturbation theory.
	
	{
		Let us also compare our results with the approach based on multipole noises \cite{Pechen2002, Pechen2004}. As this approach is trying to obtain asymptotic dynamics for the reservoir too, it is based on multipole expansion of its correlation function, which takes the form
		\begin{equation}\label{eq:trueAssOfCorrFun}
			\frac{1}{\lambda^2} G\left(\frac{t}{\lambda^2}\right)  = \tilde{G}_0 \delta(t) + \lambda^2 \tilde{G}_1 \delta'(t) + O(\lambda^4)
		\end{equation}
		in our notation. In our approach we are interested in the asymptotic expansion for the reduced dynamics only. Simple formula \eref{eq:xPert} for $ x(t;\lambda)|_{\rm pert} $ allows one to interpret $ x(t;\lambda)|_{\rm pert} $ as a solution of \eref{eq:mainIntDiff} with 
		\begin{equation}
			G_{\rm eff}(t) = - \tilde{p}(\lambda) \delta(t)
		\end{equation}
		instead of $ G(t) $ (and with initial condition $ r(\lambda) $ instead of $ 1 $). $ G_{\rm eff}(t) $ differs from true asymptotic  expansion \eref{eq:trueAssOfCorrFun} of the reservoir correlation function (except { for the zeroth} order of perturbation theory). But they lead to the same asymptotic behavior of $  x(t;\lambda)|_{\rm pert} $ and, hence, are equivalent if one is interested only in evolution of the system. If one considered infinite order multiple expansion, one would obtain the derivatives of { the} delta-function of arbitrary order in \eref{eq:trueAssOfCorrFun} and, hence, the differential equations of infinite order after naive substitution of \eref{eq:trueAssOfCorrFun} into \eref{eq:mainIntDiff}. But if one uses $ G_{\rm eff}(t) $, one obtains the first order differential equation instead. Thus, we think our approach could be more simple than the one based on the multipole noises in the case, when we are interested in the reduced dynamics only.
	}
	
	\section{Size of non-physical initial layer}
	\label{sec:cond}
	
	If $ \Re \;\tilde{G}_1 < 0 $, then initial condition \eref{eq:initCondPert} is non-physical. Here, by physical we mean that $ \rho(t;\lambda)|_{\rm pert} $ is a density matrix (hermitian, non-negative and with trace equal to 1). The estimate time needed for $ \rho(t;\lambda)|_{\rm pert} $ to become physical  is (see \ref{app:cond}) 
	\begin{equation}\label{eq:tStar}
		t^* \simeq -\lambda^2 \frac{\Re \tilde{G}_1}{\Re \tilde{G}_0}
	\end{equation}
	if $ \Re \tilde{G}_0 \neq 0 $. To understand  which properties of the reservoir spectral density contribute to $ 	t^* $, let us consider such a spectral density which can be approximated by { a} finite sum of Lorentz peaks  
	\begin{equation}
		J(\omega) =  \sum_{l=1}^n\frac{2 \gamma_l g_l^2}{ \gamma_l^2 + (\omega -\omega_l)^2}, \qquad g_l > 0, \gamma_l >0.
	\end{equation}
	Actually there is a well-developed theory of pseudomodes for such a spectral density, which was initially developed exactly for this model \cite{Imamoglu94, Garraway96, Garraway97, Garraway97a, Dalton01, Garraway06, Teretenkov19m} but then was generalized to other models  \cite{Schonleber15, Tamascelli18, Mascherpa19, Tamascelli19, Pleasance20}. The spectral density is related to the bath correlation function  $ G(t) = \frac{1}{2 \pi} \int e^{-i (\omega - \Omega) t} J(\omega)$  
	(we follow \cite{Garraway97} in the normalization constant choice). So we have
	\begin{equation}\label{eq:combOfLorPeaks}
		G(t) = \sum_{l=1}^n g_l^2 e^{-(\gamma_l + i \Delta \omega_l) t }, \qquad t \geq 0,  
	\end{equation}
	where $ \Delta \omega_l \equiv \omega_l - \Omega $. Then we obtain (see \ref{app:Comb})
	\begin{equation}\label{eq:tStarLorPeaks}
		t^* \simeq \lambda^2 \sum_{l=1}^n \frac{J_l(\Omega)}{J(\Omega)} \frac{1}{\gamma_l} \frac{\gamma_l^2 -  \Delta \omega_l^2}{\gamma_l^2 +  \Delta \omega_l^2},
	\end{equation}
	where $ J_l(\omega) = \frac{2 \gamma_l g_l^2}{ \gamma_l^2 + (\omega -\omega_l)^2} $ are  spectral densities of individual peaks.
	Hence, the absolute values $ |\Delta \omega_l| $   of detuning between the Lorentz peaks and $ \Omega $ should be high enough for absence of a non-physical initial layer. So the  non-physical initial layer occurs, when the most intense peaks of the spectral density are close to resonance.
	
	For $ l=1 $ we have
	\begin{equation}
		t^* \simeq \lambda^2 \frac{1}{\gamma_1} \frac{\gamma_1^2 -  \Delta \omega_1^2}{\gamma_1^2 +  \Delta \omega_1^2}.
	\end{equation}
	The factor $ \gamma_1^{-1} $ means that the size of the non-physical initial layer is less than the bath correlation time.
	
	\section{Weak coupling at short times and overlap}
	\label{sec:uniform}
	
	For short time $ t = O(\lambda^2) $, i.e. of order of the bath correlation time, the asymptotic expansion could be obtained just by the Dyson series for \eref{eq:mainIntDiff}  (see \ref{app:shortTime} for the expansion of arbitrary order):
	\begin{equation}\label{eq:shortTimeExpSec}
		x(t; \lambda)|_{\rm corr} = 1 - \int_0^{t} d\tau \int_0^{\tau} ds \frac{1}{\lambda^2} G\left(\frac{\tau-s}{\lambda^2}\right) + O(\lambda^4).
	\end{equation}
	
	It is possible to obtain a uniform  asymptotic expansion which neglects the terms of order $ O(\lambda^4) $ both at long and short time by asymptotic matching. Namely, we have to sum the  long and short time asymptotic expansions, but subtract the overlap term which contributes to both of them 
	\begin{equation}\label{eq:uniform}
		x(t; \lambda)|_{\rm uniform} = x(t; \lambda)|_{\rm corr} + x(t; \lambda)|_{\rm pert} - x(t; \lambda)|_{\rm overlap} .
	\end{equation}
	The overlap term has the form (see \ref{app:shortTime})
	\begin{equation}\label{eq:overlap}
		x(t; \lambda)|_{\rm overlap} = 1 -\tilde{G}_0 t - \tilde{G}_1  \lambda^2 .
	\end{equation}
	Hence, the uniform asymptotic expansion
	\begin{equation}
		\eqalign{
			x(t; \lambda)|_{\rm uniform} =&   e^{-\tilde{G}_0 t} \left(1 - \tilde{G}_1 (1 - \tilde{G}_0 t) \lambda^2 \right) + \tilde{G}_0 t  + \tilde{G}_1  \lambda^2 \nonumber  \\ 
			&  -\int_0^{t} d\tau \int_0^{\tau} ds \frac{1}{\lambda^2} G\left(\frac{\tau-s}{\lambda^2}\right) + O(\lambda^4).
		}
	\end{equation}
	The uniform expansion satisfies the initial condition $ x(0; \lambda)|_{\rm uniform} = 1 $. 
	
	\section{Explicit example}
	\label{sec:explicit}
	
	\subsection{The case of one resonance Lorentz peak}
	
\begin{figure}[t]
		\includegraphics[width=\columnwidth]{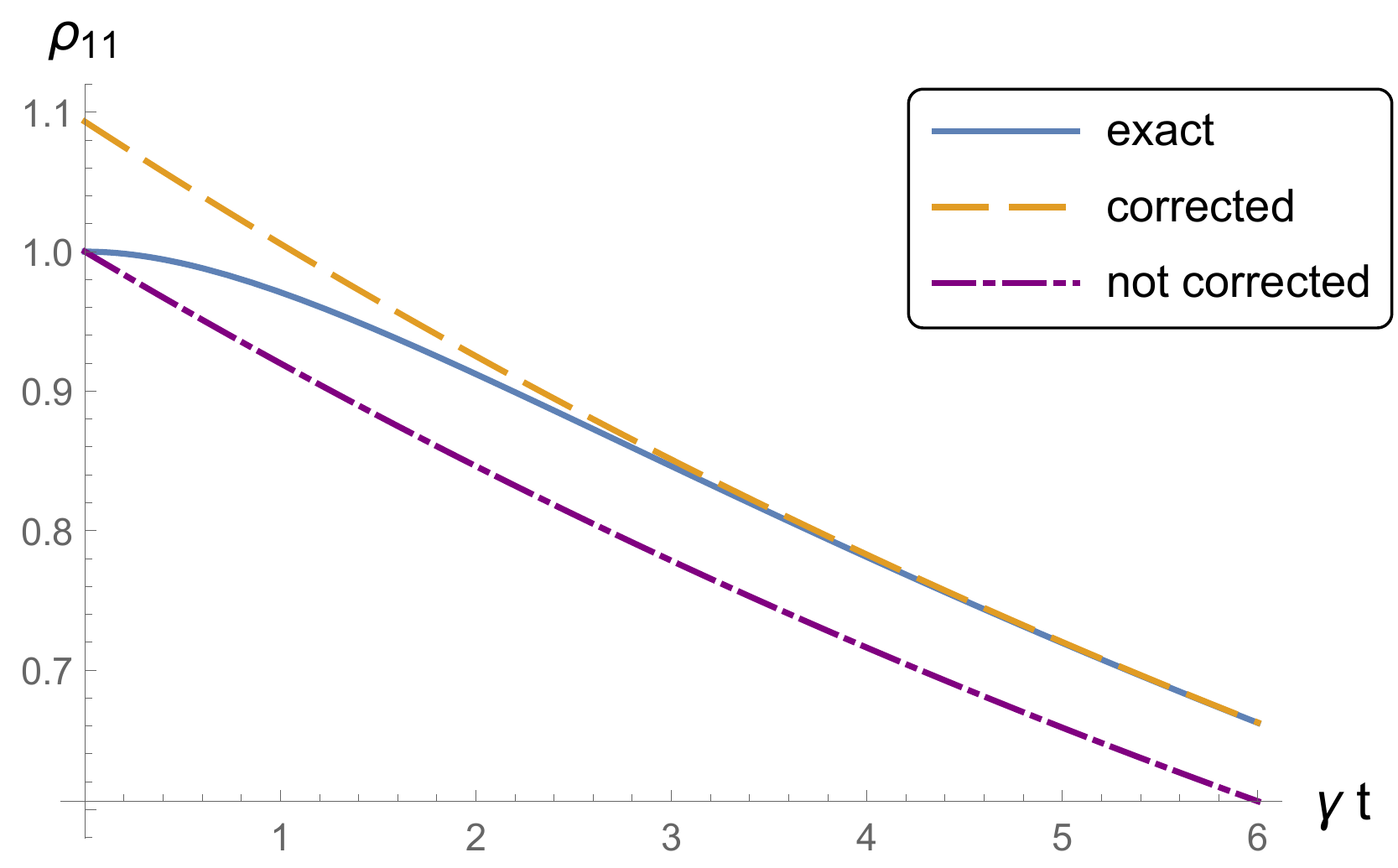}
		\caption{Dependence of the excited state population $ \rho_{11}(t) $ on time in the units of correlation time, i.e. on $ \gamma t $, with $ \rho_{11}(0) =1 $. Exact dynamics (solid line), the perturbative part with the corrected initial condition (dashed line) and the perturbative part without this correction (dot-dashed line), i.e. the one which satisfies the initial condition for the exact density matrix, are depicted. It is seen that the dynamics of { the} perturbative part with the corrected initial condition becomes indistinguishable from the exact one after $ \sim 3 $ reservoir correlation times $ \gamma^{-1} $, while dynamics of the perturbative part with the initial condition without the corrections visibly differs from the exact dynamics for the same $ t $. It { illustrates} the necessity to make such correction to reproduce a true asymptotic expansion. (The parameters are such that $ \lambda =1 $, $ g/\gamma = 0.4 $.)} 
		\label{fig_0}
	\end{figure}

	For the case when there is only one Lorentz peak with resonance, i.e. $ G(t) =  g^2 e^{- \gamma t }$, $ t \ge 0 $, the solution could be obtained explicitly \cite[subsection~10.1.2]{Breuer02}:  
	\begin{equation}
		x(t; \lambda) = \frac{1}{2\Delta} \left(\frac{\gamma}{2} + \Delta \right) e^{- \left(\frac{\gamma}{2} - \Delta\right) \frac{t}{\lambda^2}} - \frac{1}{2\Delta} \left(\frac{\gamma}{2} - \Delta \right) e^{- \left(\frac{\gamma}{2} + \Delta\right) \frac{t}{\lambda^2}}, \label{eq:expicitSol} 
	\end{equation}
	where
	\begin{equation}\label{eq:Delta}
		\Delta = \sqrt{\left(\frac{\gamma}{2}\right)^2 - \lambda^2 g^2}.
	\end{equation}
	Here, $\frac{1}{\lambda^2} \left(\frac{\gamma}{2} - \Delta\right) = \frac{g^2}{\gamma} + O(\lambda^2) $ and $\frac{1}{\lambda^2} \left(\frac{\gamma}{2} + \Delta\right) = \frac{\gamma}{2 \lambda} + O(1) $ for $ \lambda \rightarrow 0 $. So the perturbative part in this case could be calculated explicitly
	\begin{equation}
		x(t; \lambda)|_{\rm pert} =  \frac{1}{2\Delta} \left(\frac{\gamma}{2} + \Delta \right) e^{- \left(\frac{\gamma}{2} - \Delta\right) \frac{t}{\lambda^2}},
	\end{equation}
	which has form \eref{eq:xPert} with
	\begin{equation}
		r(\lambda) = \frac{1}{2\Delta} \left(\frac{\gamma}{2} + \Delta \right), \qquad p(\lambda) =  - \frac{1}{\lambda^2} \left(\frac{\gamma}{2} - \Delta\right).
	\end{equation}
	Taking into account \eref{eq:demMatEvol} one can obtain results depicted in figure~\ref{fig_0}. { The precise definition of the bath correlation time for { an} abstract correlation { function} could be debated, but for this explicit example it is consensual  \cite[subsection~10.1.2]{Breuer02} that it equals $ \gamma^{-1} $ after assuming { that} $ \lambda = 1 $. This allows us to measure time in the units of the correlation time of the plot.}
	
\begin{figure}[t]
		\includegraphics[width=\columnwidth]{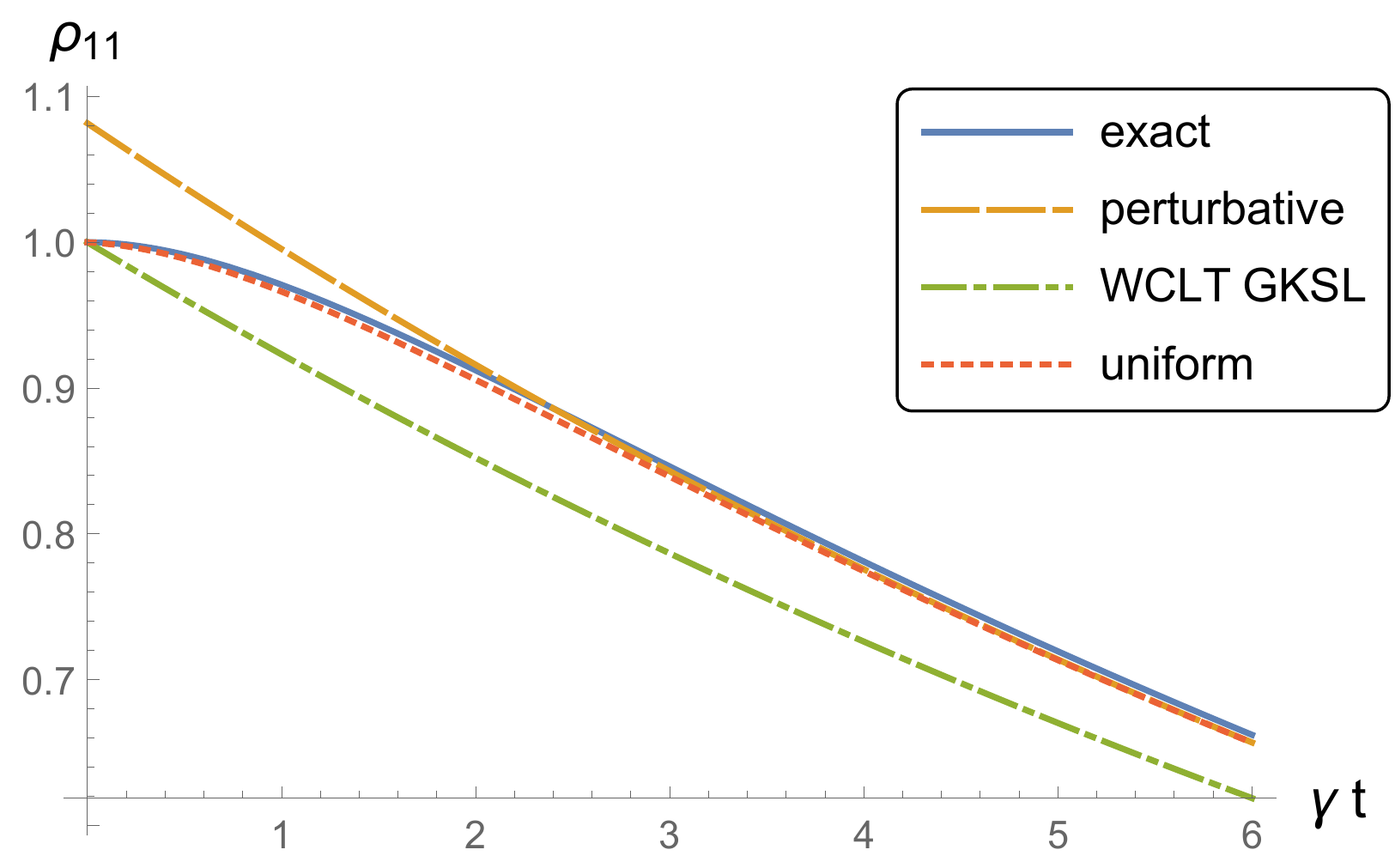}
		\caption{Dependence of the excited state population $ \rho_{11}(t) $ on time in the units of correlation time, i.e. on $ \gamma t $, with $ \rho_{11}(0) =1 $.  Exact dynamics (solid line), the perturbative part within the second order asymptotic precision (dashed line), WCLT GKSL dynamics without any corrections (dot-dashed line) and uniform second order approximation (dotted line). After $ \sim 2 $ reservoir correlation times $ \gamma^{-1} $ the perturbative part within the second order asymptotic precision becomes much closer to the exact result than WCLT GKSL dynamics without any corrections. Moreover, the uniform expansion is close to the exact result for all $ t $. (The parameters are such that $ \lambda =1 $, $ g/\gamma = 0.4 $.)} 
		\label{fig_1}
	\end{figure}

	The uniform asymptotic expansion could be also done directly:
	\begin{eqnarray}
		x(t; \lambda)|_{\rm uniform} &= e^{-\frac{g^2}{\gamma} t} \left(1 +\frac{g^2}{\gamma^2} \left(1 - \frac{g^2 }{\gamma} t\right) \lambda^2 \right) - \frac{g^2}{\gamma^2} \lambda^2  e^{- \gamma \frac{t}{\lambda^2 }} + O(\lambda^4). \label{eq:uniformExplicit}
	\end{eqnarray}
	For finite $ t $ the term $ e^{- \gamma \frac{t}{\lambda^2 }} $ is non-perturbative. And we obtain
	\begin{equation}\label{eq:specialPert}
		x(t; \lambda)|_{\rm pert} = e^{-\frac{g^2}{\gamma} t} \left(1 +\frac{g^2}{\gamma^2} \left(1 - \frac{g^2 }{\gamma} t\right) \lambda^2 \right) + O(\lambda^4).
	\end{equation} 
	Taking into account \eref{eq:demMatEvol} one can obtain results depicted in figure~\ref{fig_1}.
	
	By \eref{eq:tStarLorPeaks} the size of the non-physical initial layer for the perturbative term for such precision is $ 	t^* = \frac{1}{\gamma} \lambda^2 $, i.e. equals the bath correlation time. Relaxation rate \eref{eq:likeGKSLParam} takes the form 
	\begin{equation}
		\Gamma(t) = \frac{2g^2 }{\gamma } \left(1 +  \frac{g^2}{\gamma^2}  \lambda^2  + O(\lambda^4)\right).
	\end{equation}
	The fact that the dynamics of the perturbative part of the density matrix has the GKSL form coincides with \cite[section~V]{Haikka2011}, where both  { BLP and RHP} measures of non-Markovianity are zero until $ \Delta $ is real. Zero RHP measure is equivalent to CP-divisibility, i.e. positivity of $ \Gamma(t) $ \cite{Hall2014}, but here we prove that it is not only positive but also time-independent if only the perturbative part is considered. It is interesting that in this case zero RHP and BLP also correspond  to convergence of perturbation series for the correlation function $ \langle \sigma_-(t_2) \sigma_+(t_1)\rangle $. Moreover, this correlation function as discussed at the end of section~\ref{sec:assympInfDiv} coincides with the Markovian one multiplied by the fixed constant. Hence, the non-Markovianity measure from \cite{Gullo2014} based on the ratio of Markovian and exact correlation functions is not zero, but if we renormalize, then it is zero. Thus, in spite of the fact that it is not Markovian in the sense of the correlation function, after this renormalization it is Markovian in all the range when it is Markovian according to BLP and RHP measures. 
	
	Let us stress that { the} long-time Markovian behavior of the perturbative part does not mean that long-time dynamics is  Markovian for an arbitrary coupling constant. Due to square root in  \eref{eq:Delta} for $ \Delta $, the series for the perturbative part has a finite radius of convergence.  It rather means stability of standard Markovian approximation which occurs in the limit $ \lambda \rightarrow 0 $, i.e. such a Markovian behavior persists until the coupling constant becomes too strong. 
	
	For short times $ t = O(\lambda^2) $, we obtain
	\begin{equation}\label{eq:speciaCorr}
		x(t; \lambda)|_{\rm corr} = 1 - \frac{g^2}{\gamma} t + \frac{g^2 \lambda^2}{\gamma^2} \left(1 - e^{- \gamma \frac{t}{\lambda^2}}\right)  + O(\lambda^4).
	\end{equation}
	The overlap term could be obtained both by omitting $ e^{- \gamma \frac{t}{\lambda^2}} $ at finite $ t $ or by expanding $ x(t; \lambda)|_{\rm pert} $ for $ t = O(\lambda^2) $, which leads to
	\begin{equation}
		x(t; \lambda)|_{\rm overlap} = 1 - \frac{g^2}{\gamma} t + \frac{g^2}{\gamma^2}  \lambda^2  + O(\lambda^4)
	\end{equation}
	and also coincides with general formula \eref{eq:overlap}. By \eref{eq:uniform} we recover \eref{eq:uniformExplicit}.

	\subsection{Asymptotic precision of integro-differential and time-convolutionless equations}
	
	Let us compare this result with asymptotic behavior of the solution of the integro-differential equation in { the} Born approximation and the time-convolutionless master equation. The solution of the  integro-differential equation in { the} Born approximation has the form 
	\begin{equation}
		\rho_{B}(t) = 
		\left(\begin{array}{cc}
			x'(t) \rho_{11}(0) & x(t) \rho_{10}(0)\\
			x^*(t) \rho_{01}(0) & \rho_{00}(0) + (1-x'(t) )\rho_{11}(0)
		\end{array}\right),
	\end{equation}
	where $ x(t) $ is defined by \eref{eq:expicitSol} and $ x'(t) $ has the form similar to \eref{eq:expicitSol}, but $ \Delta $ defined by \eref{eq:Delta} should be replaced by $ \Delta' =  \sqrt{\left(\frac{\gamma}{2}\right)^2 - 2 \lambda^2 g^2} $ \cite[Equation~10.51]{Breuer02}. Hence, the coherences are predicted by this equation exactly \cite{Teretenkov19}. At the same time for population of the excited state we have
	\begin{equation}
		(\rho_B)_{11}(t) = e^{-2\frac{g^2}{\gamma} t} \left(1 + 2\frac{g^2}{\gamma^2} \left(1 - 2\frac{g^2 }{\gamma} t\right) \lambda^2 \right) \rho_{11}(0)+ O(\lambda^4).
	\end{equation}
	for the Born approximation and
	\begin{equation}
		\rho_{11}(t) = e^{-2\frac{g^2}{\gamma} t} \left(1 + 2\frac{g^2}{\gamma^2} \left(1 - \frac{g^2 }{\gamma} t\right) \lambda^2 \right) \rho_{11}(0)  + O(\lambda^4)
	\end{equation}
	for the exact solution. These expressions do not coincide already for $ O(\lambda^2) $.	Thus, for long times the integro-differential equation in the Born approximation does not guarantee the precision of $ O(\lambda^2) $ terms. For short-time expansion $ t= O(\lambda^2) $ these expansions asymptotically coincide: $ \rho_{11}(t)= (\rho_B)_{11}(t) + O(\lambda^4)  $, where
	\begin{equation}
		\rho_{11}(t)= \left(1 + \frac{2 g^2}{\gamma^2}\left(\lambda^2 \left(1 - e^{- \gamma \frac{t}{\lambda^2 }}\right) - \gamma t\right)\right)\rho_{11}(0) + O(\lambda^4).
	\end{equation}
	Hence, the integro-differential master equation is valid only on short times and its long-time precision is uncontrollable. It is exact for coherences and valid for populations only in the zeroth order of perturbation theory, i.e.  for populations it is not better than the GKSL equation with constant coefficients.

	The time-convolutionless master equation of the second order has form \eref{eq:likeGKSL} with $ \Delta \Omega_{\rm TCL2}(t) =0 $ and \cite[p.~469, (10.53)]{Breuer02}
	\begin{equation}
		\Gamma_{\rm TCL2}(t) = \frac{2g^2}{\gamma} \left( 1 - e^{-\gamma \frac{t}{\lambda^2} }\right).
	\end{equation}
	Hence, its solution has form \eref{eq:demMatEvol} with $ x_{\rm TCL2}(t) = \exp \left(-\frac12 \int_0^t \Gamma_{\rm TCL2}(\tau) d\tau\right)$. For long time we obtain
	\begin{equation}
		x_{\rm TCL2}(t) = e^{-\frac{g^2}{\gamma} t} \left(1 + \frac{g^2}{\gamma^2} \lambda^2 \right) + O(\lambda^4).
	\end{equation}
	Therefore it also differs from \eref{eq:specialPert} already in terms of order $ O(\lambda^2) $ and so do the correspondent populations and coherences. For short time $ t= O(\lambda^2) $ it coincides with \eref{eq:speciaCorr}.
	
	So, to take into account the long-{ time} behavior, further corrections have to be taken into account. For example, the time-convolutionless master equation of the fourth order has form \eref{eq:likeGKSL} with $ \Delta \Omega_{\rm TCL4}(t) =0 $ and 
	\begin{equation}
		\Gamma_{\rm TCL4}(t) = \frac{2g^2}{\gamma} \left( 1 - e^{- \frac{\gamma}{\lambda^2} t}\right)+ \lambda^2 \frac{4 g^4}{\gamma^3} e^{- \frac{\gamma}{\lambda^2} t} \left(\sinh \left(\frac{\gamma}{\lambda^2} t\right) - \frac{\gamma}{\lambda^2} t\right)
	\end{equation}
	in this case \cite[p.~469, Equations 10.53--10.54]{Breuer02}. Once again calculating $ x_{\rm TCL4}(t) = \exp \left(-\frac12 \int_0^t \Gamma_{\rm TCL4}(\tau) d\tau\right)$ we obtain that its asymptotic expansion now coincides with \eref{eq:specialPert} for fixed $ t $. 
	
	Hence, the time-convolutionless master equation gives non-uniform asymptotic expansion which is better for short times than for long times, but higher precision of the long-time asymptotic expansion could be achieved by the higher order time-convolutionless master equation. It could be understood in the general case from the fact that the derivative is multiplied by $ \lambda^2 $ after { the} scaling  $ t \rightarrow \lambda^{-2} t  $. 
	
	\section{Conclusions}
	
	For RWA spin-boson at zero temperature we have derived perturbative correction to the WCLT GKSL equation which arises in the Bogolubov-van Hove limit of this model.   We have shown that the initial conditions for the whole density matrix and its perturbative part do not coincide. Moreover, in certain cases they have to be  non-physical to reproduce true asymptotic expansion if only perturbative terms are included.
	
	We have also shown that the perturbative part satisfies a certain { master equation with the GKSL generator} and the most important result is that this generator is time-independent. The perturbative two-time correlation function considered in section~\ref{sec:assympInfDiv} coincides with the Markovian expression multiplied by the time-independent factor which also arises from this change of initial conditions.  So all non-Markovian effects occur on the bath correlation time and manifest themselves at long times only in the change of initial conditions.  That is why we have called such a behavior a long-time Markovian one. But let us emphasize that such { a} behavior of system correlation functions is not Markovian in the sense of \cite[section~3.4]{Li2018} as discussed in the second paragraph of section~\ref{sec:assympInfDiv} and, hence, in other weaker senses, e.g. the past-future independence is broken here in the strict sense (see \cite[subsection~4.1.2]{Li2018}). { So, as we have discussed in the Introduction{,} our long-time Markovianity could be thought as the intermediate regime between Markovianity and non-Markovianity.}
	
	Moreover, the localization of non-Markovian effects inside the region of order of correlation time is valid in all the orders of perturbation theory only in the case of the reservoir correlation functions with all finite moments. In general, it is violated in the order determined by first infinite moment of the reservoir correlation function. In \ref{app:nonUniversality} we  consider the example which shows that the long-time Markovianity even in the second order of perturbation theory in $ \lambda $ is not generic.
	
	The non-physical behavior of the perturbative part of the reduced density matrix is localized at bath correlation time and occurs only under certain resonance conditions discussed in section~\ref{sec:cond}. By matching with { the} short-time expansion the uniform expansion was obtained in section~\ref{sec:uniform}. All our discussion was illustrated by the example in section~\ref{sec:uniform}. In particular, we have shown that our master equations differ from the standard time-convolutionless master equation, but have the same asymptotic precision if the small parameter is introduced as in the Bogolubov-van Hove limit.
	
	So the perturbation theory with the Bogolubov-van Hove scaling  highlights the physical behavior which is not so easily seen in other approaches. { Hence, our definition of long-time { Markovianity} along with the perturbative analysis based on  the Bogolubov-van Hove scaling could be useful for other physical models. Actually, the generalization of the results discussed here on RWA models such as the ones from \cite{Garraway97} or \cite{Teretenkov19} seems to be straightforward as equation (12) from \cite{Teretenkov19} has the structure similar to \eref{eq:intDiffIntPic}. The possibility to write down time-independent asymptomatic master equations also seems to be generalizable even for non-RWA models. For this case one should apply the asymptotic analysis developed in this paper directly to the Nakajima-Zwanzig equation. But it is not obvious if they have the GKSL form or if the correlation functions could be defined by the renormalized Markovian formulae. So the main  direction for further study is to clarify these questions for more general models.}

	\ack The author expresses his gratitude to  A.\,S.~Trushechkin for the fruitful discussion of the problems considered in the work and valuable remarks on its text. The author thanks I.\,Sinayskiy and I.\,V.~Volovich for important bibliographic { references. The author is very grateful to the Referees for the important remarks.}  This work is supported by the Russian Science Foundation under grant 17-71-20154.
	
	\appendix
	
	\section{Expansion}
	\label{app:Exp}
	
	Let us apply the Laplace transform to \eref{eq:mainIntDiff}, then
	\begin{equation}
		p \tilde{x}(p; \lambda) - 1= -\tilde{G}(\lambda^2 p) \tilde{x}(p; \lambda),
	\end{equation}
	where $ \tilde{G}(p) \equiv \int_0^{+\infty} e^{-pt}  G(t) dt $ and $ \tilde{x}(p; \lambda) \equiv \int_0^{+\infty} e^{-pt}  x(t; \lambda) dt$, which is solved as
	\begin{equation}
		\tilde{x}(p; \lambda) = \frac{1}{p + \tilde{G}(\lambda^2 p) }.
	\end{equation}
	Now, let us expand  $ \tilde{G}(p) $ in the Taylor series
	\begin{equation}\label{eq:GTaylorSeries}
		\tilde{G}(p) = \sum_{k=0}^{\infty}  \tilde{G}_{k} p^k,
	\end{equation}
	where
	\begin{equation}\label{eq:mometOfG}
		\tilde{G}_k = \frac{(-1)^k}{k!}\int_0^{\infty} t^k  G (t) dt.
	\end{equation}
	
	Then the series expansion in $ \lambda $ for $ \tilde{x}(p;\lambda)  $ has the form
	\begin{equation}
		\tilde{x}(p;\lambda) = \sum_{k=0}^{\infty}\lambda^{2k} p^k \tilde{x}_k(p),
	\end{equation}
	where the coefficients could be obtained by the Wronski formula \cite[p.~17]{Henrici74}:
	\begin{equation}
		\tilde{x}_k(p) = \frac{(-1)^k p^k}{(p+\tilde{G}_0)^{k+1}} D_k(p),
	\end{equation}
	where
	\begin{equation}
		D_k(p)= \det
		\left(\begin{array}{ccccc}
			\tilde{G}_1 & \tilde{G}_2  & \dots & \dots & \tilde{G}_k \\
			\tilde{G}_0 + p & \tilde{G}_1  & \dots &  \dots & \tilde{G}_{k-1} \\
			0 & \tilde{G}_0 + p &  \ddots & \dots &  \tilde{G}_{k-2} \\
			\vdots & 0 & \ddots & \ddots & \vdots\\
			0 & 0 & \dots &  \tilde{G}_0 + p & \tilde{G}_1 
		\end{array}\right).
	\end{equation}
	The explicit formula for $ k=0,1,2 $:
	\begin{equation}
		\eqalign{
			\tilde{x}_0(p) &= \frac{1}{p+\tilde{G}_0}, \nonumber\\
			\tilde{x}_1(p) &=- p \frac{\tilde{G}_1 }{(p+\tilde{G}_0)^2}, \nonumber\\
			\tilde{x}_2(p) &= p^2\frac{(\tilde{G}_1 )^2 - (p+\tilde{G}_0)\tilde{G}_2 }{(p+\tilde{G}_0)^3}.}
	\end{equation}
	Inverting the Laplace transform we obtain $ t>0 $
	\begin{equation}
		x(t; \lambda)|_{\rm pert} = \sum_{k=0}^{\infty} x_k(t) \lambda^{2k},
	\end{equation}
	where
	\begin{equation}\label{eq:expanTimeDom}
		x_k(t) = \frac{(-1)^k}{k!}
		D_k \left(\frac{d}{dt}\right) \frac{d^k}{dt^k} (t^{k} e^{- \tilde{G}_0 t}).
	\end{equation}
	In particular, this formula means that the term $ x_k(t) $ is a polynomial of the degree less or equal to $ k $ times $  e^{- \tilde{G}_0 t} $.
	
	The explicit formula for $ k=0,1,2 $:
	\begin{equation}
		\eqalign{
			x_0(t) &= e^{-\tilde{G}_0 t}, \\
			x_1(t) &= -\tilde{G}_1 (1 - \tilde{G}_0 t) e^{-\tilde{G}_0 t}, \\
			x_2(t) &= \left(\tilde{G}_1^2 + 2 \tilde{G}_0 \tilde{G}_2 - \tilde{G}_0 (2 \tilde{G}_1^2 + \tilde{G}_0 \tilde{G}_2) t + \frac12 \tilde{G}_0^2 \tilde{G}_1^2 t^2\right) e^{-\tilde{G}_0 t}.}
	\end{equation}
	Taking into account only terms for $ k=0,1 $ we obtain \eref{eq:xSecOrder}.
	
	Let us mention that if the $ n $-th moment of $ G (t) $ does not exist (but all { the} lower orders do), then one could obtain only an expansion to the finite order
	\begin{equation}
		\tilde{x}(p;\lambda) = \sum_{k=0}^{n-1}\lambda^{2k} p^k \tilde{x}_k(p) + O(\lambda^{2n}).
	\end{equation}
	
	\section{Perturbative part in exponential form}
	\label{app:pertPartDiffEq}
	
	From \eref{eq:GTaylorSeries} we have
	\begin{equation}
		\tilde{G}(\lambda^2 p) = \sum_{k=0}^{n}  \tilde{G}_{k} \lambda^{2k} p^k + O(\lambda^{2n+2}),
	\end{equation}
	then
	\begin{equation}
		\frac{1}{p + \tilde{G}(\lambda^2p)} = \frac{1}{p+\sum_{k=0}^{n}  \tilde{G}_{k} \lambda^{2k} p^k} + O(\lambda^{2n+2}),
	\end{equation}
	i.e. we have a polynomial in the denominator. If we assume a generic form of $ \tilde{G}(p) $, then $ \tilde{G}_{n} \neq 0 $ and this polynomial can be factorized as
	\begin{equation}
		p+\sum_{k=0}^{n}  \tilde{G}_{k} \lambda^{2k} p^k = (p - \tilde{p}) \lambda^{2n} \tilde{G}_{n} \prod_{k=1}^{n-1} \left(p - \frac{p_k}{\lambda^2}\right),
	\end{equation}
	where $ \tilde{p} = O(1) $ for $ \tilde{G}_{0} \neq 0 $ (for  $ \tilde{G}_{0} = 0 $ $ \tilde{p} = o(1) $),  $ p_k  = O(1) $, which can be found by the Newton diagram method \cite{Vainberg1974, White2010}. In the generic case we also can assume that there are no coinciding $ \tilde{p}_k $
	\begin{equation}\label{eq:assympPolesExp}
		\tilde{x}(p; \lambda) = \frac{r(\lambda)}{p - \tilde{p}(\lambda)} + \sum_{k=1}^{n-1} \frac{r_k(\lambda)}{p -\frac{ p_k(\lambda)}{\lambda^2}} + O(\lambda^{2n+2}).
	\end{equation}
	After the inverse Laplace transform we obtain
	\begin{equation}
		x(t; \lambda) = r(\lambda) e^{\tilde{p}(\lambda) t} + \sum_{k=1}^{n-1} r_k(\lambda) e^{p_k(\lambda) \frac{t}{\lambda^2}} + O(\lambda^{2n+2}).
	\end{equation}
	Hence, the perturbative part for any $ n $ has the form $ x(t; \lambda)|_{\rm pert} = r(\lambda) e^{\tilde{p}(\lambda) t} $. If $  \tilde{G}_{n} = 0  $, then it means just that there would be fewer terms $  r_k(\lambda) e^{p_k(\lambda) \frac{t}{\lambda^2}} $ with the same asymptotic behavior of $ p_k(\lambda) $. If now some of $ p_k(\lambda) $ coincide, then terms of the form $ \frac{r_k(\lambda)}{(p - \lambda^{-2} p_k(\lambda))^{m+1}}  $ may occur in \eref{eq:assympPolesExp}. After the inverse Laplace transform they have the form $ r_k(\lambda) \frac{t^m}{m!} e^{p_k(\lambda) \frac{t}{\lambda^2}} $, so they are also non-perturbative. Due to the condition that $ |x(t; \lambda)|^2 \leq 1  $ which follows from the positivity of the exact density matrix we have $ \Re p_k(\lambda) \leq 0 $, because otherwise the condition $ |x(t; \lambda)|^2 \leq 1  $ would be violated more than the error $ O(\lambda^{2n+2}) $ allows. Certain caution is needed in the case when $ \Re p_k(\lambda) = 0 $ for some $ k $ and $ n $. These terms also do not contribute to the perturbative part, but they are rapidly oscillating rather than rapidly decaying at long times. So they do not vanish, but could be neglected if only the average behavior { were} observable.
	
	So let us calculate the Laplace transform of the perturbative part
	\begin{equation}
		\tilde{x}(p; \lambda)|_{\rm pert} = \frac{r(\lambda)}{p - \tilde{p}(\lambda)}.
	\end{equation}
	$ \tilde{p}(\lambda) $ could be found as the solution of the equation
	\begin{equation}\label{eq:pole}
		\tilde{p}(\lambda) + \tilde{G}(\lambda^2 \tilde{p}(\lambda)) = 0
	\end{equation}
	by the perturbative series $ \tilde{p} = \sum_{n=0}^{\infty} \tilde{p}_n \lambda^{2n} $. Calculating the series for composition of functions \cite[(5.10)]{Stanley2001} we obtain  $ \tilde{G}(\lambda^2 \tilde{p}) = \sum_{n=0}^{\infty} \lambda^{2n} \sum_{C_n} \tilde{G}_k \tilde{p}_{i_1 -1} \ldots \tilde{p}_{i_k -1} $, where $ C_n $ are all possible compositions of $ n $, i.e. sets of integers $ i_1 \geq 1, \cdots, i_k \geq 1 $  
	such that $ i_1 + \ldots i_k = n $. Equating the asymptotic expansions for both sides of \eref{eq:pole} we obtain the recurrence equation for the $ n $-th term of expansion of $ \tilde{p}(\lambda) $
	\begin{equation}
		\tilde{p}_{n} = - \sum_{C_n} \tilde{G}_k \tilde{p}_{i_1 -1} \ldots \tilde{p}_{i_k -1}.
	\end{equation}
	The explicit form of several first terms
	\begin{equation}
		\eqalign{
			\tilde{p}_{0} &= -\tilde{G}_0, \nonumber\\
			\tilde{p}_{1} &= \tilde{G}_0 \tilde{G}_1, \nonumber\\
			\tilde{p}_{2} &= -\tilde{G}_0 (\tilde{G}_1^2 + \tilde{G}_0\tilde{G}_2), \nonumber\\
			\tilde{p}_{3} &= \tilde{G}_0 (\tilde{G}_1^3 + 3\tilde{G}_0 \tilde{G}_1\tilde{G}_2 + \tilde{G}_0^2 \tilde{G}_3).}
	\end{equation}
	$ r(\lambda) $ could be found as
	\begin{eqnarray}
		r(\lambda) = \lim\limits_{\varepsilon \rightarrow 0}  \left. \frac{\varepsilon r(\lambda)}{p - \tilde{p}(\lambda)} \right|_{p = \tilde{p}(\lambda) + \varepsilon} &=\lim\limits_{\varepsilon \rightarrow 0}  \left.\frac{\varepsilon}{p + \tilde{G}(\lambda^2 p)} \right|_{p = \tilde{p}(\lambda) + \varepsilon}  \nonumber\\
		&= \frac{1}{1 + \lambda^2 \frac{d}{dp}\tilde{G}|_{p=\lambda^2 \tilde{p}}}. \label{eq:coeff}
	\end{eqnarray}
	By differentiating \eref{eq:pole} we obtain
	\begin{equation}
		\frac{d\tilde{p}}{d \lambda^2} +\frac{d}{dp}\tilde{G}|_{p=\lambda^2 \tilde{p}} \left(\tilde{p} + \lambda^2\frac{d\tilde{p}}{d \lambda^2}\right) = 0.
	\end{equation}
	Substituting it to \eref{eq:coeff} we obtain 
	\begin{equation}
		r(\lambda) = 1 +\lambda^2\frac{1}{\tilde{p}}\frac{d \tilde{p}}{d \lambda^2} .
	\end{equation}
	The asymptotic expansion $ \lambda^2\frac{d \tilde{p}}{d \lambda^2} = \sum_{n=1}^{\infty} n \tilde{p}_{n} \lambda^{2n} $. Then we obtain $ r(\lambda) = \sum_{n=1}^{\infty} r_n \lambda^{2n} $, where $ r_0 = 1 $ and
	\begin{equation}
		r_n = \frac{1}{\tilde{p}_{0}} \left(n\tilde{p}_{n}-\sum_{k=1}^{n} \tilde{p}_{k} (n-k) r_{n-k}\right).
	\end{equation}
	The explicit formulae for several first terms 
	\begin{equation}
		\eqalign{
			r_1 &= - \tilde{G}_1, \nonumber\\
			r_2 &= \tilde{G}_1^2 + 2 \tilde{G}_0\tilde{G}_2, \nonumber\\
			r_3 &= - (\tilde{G}_1^3 + 6 \tilde{G}_0 \tilde{G}_1 \tilde{G}_2 + 3 \tilde{G}_0^2 \tilde{G}_3).}
	\end{equation}
	
	\section{Correlation function}
	\label{app:corrFun}
	
	Let us denote the evolution of the system and reservoir in the interaction picture by $ U_t $. Then \cite{Teretenkov19m}
	\begin{equation}
		\eqalign{
			U_{t} | 0 \rangle \otimes | {\rm vac} \rangle &= | 0 \rangle \otimes | {\rm vac} \rangle, \nonumber\\
			U_{t} | 1 \rangle \otimes | {\rm vac} \rangle &= x(t)  | 1 \rangle \otimes | {\rm vac} \rangle + \int dk \psi_k(t) | 0 \rangle \otimes b_k^{\dagger} | {\rm vac} \rangle,}
	\end{equation}
	where $ x(t) $ is the solution of \eref{eq:mainIntDiff}. By direct calculation we have
	\begin{eqnarray}
		\langle \sigma_-(t_2) \sigma_+(t_1)\rangle &\equiv \Tr U_{t_2}^{\dagger} |0\rangle\langle 1| U_{t_2}  U_{t_1}^{\dagger} |1\rangle\langle 0| U_{t_1} |0\rangle\langle 0| \otimes  |{\rm vac} \rangle\langle {\rm vac} | \nonumber \\
		& =  \Tr U_{t_2}^{\dagger} |0\rangle\langle 1| U_{t_2 - t_1}   |1\rangle\langle 0| \otimes  |{\rm vac} \rangle\langle {\rm vac} | \nonumber \\
		&=   x(t_2-t_1) \Tr U_{t_2}^{\dagger} |0\rangle \langle 0| \otimes  |{\rm vac} \rangle\langle {\rm vac} | = x(t_2-t_1).
	\end{eqnarray}
	
	\section{Conditions for physical initial behavior}
	\label{app:cond}
	
	To be physical, i.e. to be a density matrix, the matrix $  \rho(t; \lambda)|_{\rm pert} $ should satisfy $ \rho_{11}(t; \lambda)|_{\rm pert} \leq 1 $ and $ \det \rho(t; \lambda)|_{\rm pert} \geq 0 $ as trace preservation and self-adjointness are provided by form \eref{eq:demMatEvol}. By direct calculation $  \det \rho(t; \lambda)|_{\rm pert} \geq 0 $ takes the form	
	\begin{equation}\label{eq:physCond}
		\biggl|x(t; \lambda)|_{\rm pert} \biggr|^2 \rho_{11}(0) \leq 1- \frac{|\rho_{10}(0)|^2}{\rho_{11}(0)} 
	\end{equation}
	for $ \rho_{11}(0) \neq 0 $, which automatically leads to  
	\begin{equation}
		\rho_{11}(t; \lambda)|_{\rm pert}= \biggl|x(t; \lambda)|_{\rm pert} \biggr|^2  \rho_{11}(0)  \leq 1.
	\end{equation}
	So only inequality~\eref{eq:physCond} should be justified for $  \rho(t; \lambda)|_{\rm pert} $ to be physical if  $ \rho_{11}(0) \neq 0 $. If  $ \rho_{11}(0) = 0  $, then $ \rho_{10}(0) = \rho_{01}(0) = 0 $ and $ \rho_{00}(0) = 1  $, which is preserved by dynamics \eref{eq:demMatEvol}. Taking into account \eref{eq:xPert}, inequality~\eref{eq:physCond} takes the form
	\begin{equation}
		|r(\lambda)| e^{ \Re \tilde{p}(\lambda) t}  \leq \frac{\sqrt{\rho_{11}(0) - |\rho_{10}(0)|^2}}{\rho_{11}(0)}.
	\end{equation}
	As $ \det \rho(0) \geq 0 $ leads to $ \rho_{11}(0) - |\rho_{10}(0)|^2 \geq ( \rho_{11}(0) )^2 $ and as $ \Re \tilde{p}(\lambda) \leq 0 $, then $  \rho(t; \lambda)|_{\rm pert} $ is physical for all times if $ |r(\lambda)| \leq 1 $ or is initially non-physical, but becomes physical at the time $ t^* $ defined by $ 	|r(\lambda)| e^{ \Re \tilde{p}(\lambda) t^*} = 1 $ if $  \Re \tilde{p}(\lambda) \neq 0  $, i.e. at
	\begin{equation}
		t^* = - \frac{\ln |r (\lambda)|}{\Re \tilde{p}(\lambda)}.
	\end{equation}
	Substituting asymptotic expansion for $ r (\lambda) $ and $ \tilde{p}(\lambda) $ from \ref{app:pertPartDiffEq} we have \eref{eq:tStar}.
	
	\section{Combination of Lorentz peaks}
	\label{app:Comb}
	
	For the correlation function of form \eref{eq:combOfLorPeaks} we have 
	\begin{equation}
		\tilde{G}(p) = \sum_{l=1}^n g_l^2  \frac{1}{p + \gamma_l + i \Delta \omega_l}.
	\end{equation}
	Then by \eref{eq:mometOfG} for $ k=0,1 $ we obtain 
	\begin{equation}
		\eqalign{
			\tilde{G}_0 &= \sum_{l=1}^n g_l^2   \frac{\gamma_l - i \Delta \omega_l}{\gamma_l^2 +  \Delta \omega_l^2} = \sum_{l=1}^n \frac{J_l(\Omega)}{2}\left(1 - i \frac{\Delta \omega_l}{\gamma_l}\right), \nonumber\\
			-\tilde{G}_1 &=  \sum_{l=1}^n g_l^2 \frac{\gamma_l^2 -  \Delta \omega_l^2 - 2 i \gamma_l  \Delta \omega_l}{\left(\gamma_l^2 +  \Delta \omega_l^2\right)^2} = \sum_{l=1}^n \frac{J_l(\Omega)}{2} \frac{\gamma_l^2 -  \Delta \omega_l^2 - 2 i \gamma_l  \Delta \omega_l}{\gamma_l (\gamma_l^2 +  \Delta \omega_l^2)}.}
	\end{equation}
	For real parts we have
	\begin{equation*}
		\Re \tilde{G}_0 =   \sum_{l=1}^n \frac{J_l(\Omega)}{2} = \frac{J(\Omega)}{2}, \quad -\Re \tilde{G}_1 = \sum_{l=1}^n \frac{J_l(\Omega)}{2} \frac{\gamma_l^2 -  \Delta \omega_l^2}{\gamma_l (\gamma_l^2 +  \Delta \omega_l^2)}.
	\end{equation*}
	Thus, by \eref{eq:tStar} we obtain \eref{eq:tStarLorPeaks}.
	
	\section{Short time expansion and matching}
	\label{app:shortTime}
	
	Let us rewrite \eref{eq:mainIntDiff} as
	\begin{equation}\label{eq:integraForm}
		x(t) =1 -  \int_0^t dt_2 \int_0^{t_2}dt_1 \frac{1}{\lambda^2} G\left(\frac{t_2 -t_1}{\lambda^2}\right) x(t_1).
	\end{equation}
	For short times $ t = O (\lambda^2)  $ and for $ y\left(\frac{\tau_1}{\lambda^2}\right) = O(\lambda^{2k}) $ we have 
	\begin{eqnarray}
		\int_0^t dt_2 \int_0^{t_2}dt_1 \frac{1}{\lambda^2} G\left(\frac{t_2 -t_1}{\lambda^2}\right) y(t_1) \nonumber \\
		= \lambda^2\int_0^{\frac{t}{\lambda^2}} d \tau_2 \int_0^{\tau_2}d\tau_1 G(\tau_2 - \tau_1) y\left(\frac{\tau_1}{\lambda^2}\right)  = O(\lambda^{2(k+1)}).
	\end{eqnarray}
	So it is possible to iterate \eref{eq:integraForm} to obtain the asymptotic expansion for   short times $ t = O (\lambda^2)  $.
	\begin{eqnarray}
		\fl
		x(t)|_{\rm corr}  =1 - \int_0^t dt_2 \int_0^{t_2}dt_1 \frac{1}{\lambda^2} G\left(\frac{t_2 -t_1}{\lambda^2}\right) \nonumber  \\
		+ \int_0^{t} dt_4 \int_0^{t_4} dt_3 \int_0^{t_3} dt_2 \int_0^{t_2}dt_1 \frac{1}{\lambda^2} G\left(\frac{t_4 -t_3}{\lambda^2}\right)  \frac{1}{\lambda^2} G\left(\frac{t_2 -t_1}{\lambda^2}\right) + \cdots
	\end{eqnarray}
	In particular, neglecting the terms  $ O(\lambda^4) $ we obtain \eref{eq:shortTimeExpSec}.
	
	To obtain the uniform asymptotic expansion, one should identify the overlap terms which contribute both to $ x(t)|_{\rm pert} $ and $ x(t)|_{\rm corr} $ and subtract them from their sum. Let us apply the Laplace transform to $ x(t)|_{\rm corr}  $. We obtain
	\begin{equation}
		\tilde{x}(p; \lambda)|_{\rm corr} =\frac{1}{p} \sum_{k=0}^{\infty} (-1)^k \left(\frac{\tilde{G}(\lambda^2 p)}{p}\right)^k.
	\end{equation}
	To obtain asymptotic expansion accurate within $ \lambda^{2n} $ terms, we need only first $ n+1 $ terms for short-time expansion
	\begin{equation}
		\tilde{x}(p; \lambda)|_{\rm corr} \simeq \frac{1}{p} \sum_{k=0}^{n} (-1)^k \left(\frac{\tilde{G}(\lambda^2 p)}{p}\right)^k.
	\end{equation}
	Now we need to identify  the terms in this approximation which  also contribute to the long-time expansion. For this purpose let us obtain expansion of $  (\tilde{G}(p))^k $ in $ p $. For the  $ k $-th power of power series for $ \tilde{G}(p) $ we have \cite[section~0.31]{Goldstein2017}
	\begin{equation}
		(\tilde{G}(p))^k = \sum_{m=0}^{\infty} \tilde{G}_{k,m} p^m,
	\end{equation}
	where
	\begin{equation}
		\tilde{G}_{k,0} = \tilde{G}_{0}^k, \qquad \tilde{G}_{k,m} = \frac{1}{m \tilde{G}_{0}}\sum_{j=1}^m(j k - m +j) \tilde{G}_{j} \tilde{G}_{k,m-j}.
	\end{equation}
	By definition $ \tilde{G}_{1, m} = \tilde{G}_{m} $. Let us present several first terms for $ k=2$ 
	\begin{equation}
		\tilde{G}_{2,0} = \tilde{G}_{0}^2, \qquad 
		\tilde{G}_{2,1} = 2 \tilde{G}_{0} \tilde{G}_{1}, \qquad
		\tilde{G}_{2,2} = \tilde{G}_{1}^2 + 2 \tilde{G}_{0} \tilde{G}_{2}.
	\end{equation}
	The only terms which contribute at long times (see \cite[section~8.4-7]{Korn2000}  or \cite[section~11.4.2]{Prosperetti11}) { come} from the negative powers of $ p $, i.e.
	\begin{equation}
		\tilde{x}(p; \lambda)|_{\rm overlap} = \sum_{k=0}^{n}\sum_{m=0}^{k} (-1)^k\tilde{G}_{k,m} \frac{p^m}{p^{k+1}} \lambda^{2m}.
	\end{equation}
	Performing the inverse Laplace transform we have
	\begin{equation}
		x(t; \lambda)|_{\rm overlap} = \sum_{k=0}^{n}\sum_{m=0}^{k} (-1)^k\tilde{G}_{k,m} \frac{t^{k-m}}{(k-m)!} \lambda^{2m}.
	\end{equation}
	Namely, we have
	\begin{equation}
		\eqalign{
			x(t; \lambda)|_{\rm overlap} = 1- \tilde{G}_0 t + \tilde{G}_1 \lambda^2 ,  & \quad  n=1,\\
			x(t; \lambda)|_{\rm overlap} = 1- \tilde{G}_0 t + \tilde{G}_1 \lambda^2 + \frac{\tilde{G}_0  t^2}{2}+ \tilde{G}_1 t \lambda^2 + \tilde{G}_2 \lambda^4 , & \quad n=2.
		}
	\end{equation}
	
	\section{Non-universality of long-time Markovian behavior in case of non-finite moments}
	\label{app:nonUniversality}

	Let us stress that the existence of finite moments function is an important non-technical assumption if one wants to obtain the asymptotic GKSL equation \eref{eq:masterEq}-\eref{eq:likeGKSL} with constant coefficients \eref{eq:likeGKSLParam}. To show it, let us consider the spectral density of the form: 
	\begin{equation}\label{eq:specDenNonDiff}
		\fl J(\omega + \Omega) = \chi g^2 \frac{2 \gamma}{\gamma^2 + \omega^2} + (1-\chi)g^2 |\omega|^{\frac12} \frac{ \sqrt{2 \gamma}}{\gamma^2 + \omega^2}, \qquad \gamma>0, g>0, \chi \in (0,1).
	\end{equation}
	Then, after the Fourier transform, we obtain the reservoir correlation function of the form
	\begin{equation}
		\fl G(t) =g^2\left(\chi e^{-\gamma |t|}+ \frac{1-\chi}{2} \left(e^{-\gamma |t|}(1- \Im \mathrm{erf}(i  \sqrt{\gamma |t|})) +  e^{\gamma|t|}(1- \mathrm{erf}\;( \sqrt{\gamma |t|}))\right)\right),
	\end{equation}
	where $ \mathrm{erf}\; $ is the error function \cite[Equation~7.2.1]{Temme2010}. It has the finite  zeroth moment $ \tilde{G}_0 = \chi \frac{g^2}{\gamma}$, but integral for $  \tilde{G}_1 $ does not converge. Its Laplace transform can be calculated explicitly:
	\begin{equation*}
		\tilde{G}(p) = g^2\left(\chi \frac{1}{p + \gamma} + (1-\chi) \frac{\sqrt{p}}{(\sqrt{p} + \sqrt{\gamma}) (p + \gamma)}\right).
	\end{equation*}
	Then by expanding 
	\begin{equation*}
		\tilde{G}( p) =  \frac{g^2}{\gamma}  \left(\chi + (1-\chi) \sqrt{\frac{p}{\gamma}} \right) + O(p)
	\end{equation*}
	we see that divergence of the integral for $ \tilde{G}_1 $ originates from the presence of non-integer powers of $ p $ in this expansion. It leads to presence of odd powers of $ \lambda $ in expansion for $ \tilde{x}(p; \lambda) $. Namely, we have
	\begin{equation}
		\tilde{x}(p; \lambda)|_{\rm pert} = \tilde{x}_0(p) + \lambda \tilde{x}_{\frac12}(p) + O(\lambda^2),
	\end{equation}
	where
	\begin{equation}
		\tilde{x}_0(p) =\frac{1}{p+\chi\frac{g^2}{\gamma} }, \qquad 	\tilde{x}_{\frac12}(p) = - \frac{g^2}{\gamma}  \frac{1 - \chi}{(p+\chi\frac{g^2}{\gamma} )^2}  \sqrt{\frac{p}{\gamma}}.
	\end{equation}
	
	Then, after the inverse Laplace transform we have
	\begin{equation*}
		x(t;\lambda^2) = x_0(t) + \lambda x_{\frac12}(t) + O(\lambda^2),
	\end{equation*}
	where
	\begin{equation}
		\fl x_0(t) = e^{- \chi \frac{g^2}{\gamma} t}, \qquad x_{\frac12}(t) = - \frac{1 - \chi}{\sqrt{\pi \chi }} \frac{g}{\gamma}\left(\sqrt{\chi \frac{g^2}{\gamma} t} + \left(1 - 2 \chi \frac{g^2}{\gamma} t\right) \mathrm{daw} \left(\sqrt{\chi \frac{g^2}{\gamma} t}\right)\right),
	\end{equation}
	$ \mathrm{daw}(x)  $ is the Dawson integral \cite[Equation~7.2.5]{Temme2010}. As $ \mathrm{daw}(x) \sim \frac{1}{2 x}, x \rightarrow + \infty $ \cite[theorem~2]{Nijimbere2019}, then  $ x_{\frac12}(t) = O(t^{-\frac12}), t \rightarrow + \infty $. Hence, we obtain non-exponential long-time behavior in the first order of perturbation theory in $ \lambda $ (one half order in $ \lambda^2$). But for the zeroth order of perturbation theory we can apply our general approach and indeed we have obtained the exponential decay. Let us note that by itself (without its asymptotic analysis) the phenomenon of the non-exponential decay of an open system is well-known \cite{Peres1980, Khalfin1958, Giraldi2015}.
	In the similar manner if the moments are finite only up to the fixed order, then single exponential formula \eref{eq:xPert} is also valid up to the fixed order of perturbation theory. \eref{eq:correlFunct} is based on \eref{eq:xPert} and, moreover, \eref{eq:MarkCorrFun} and \eref{eq:exactCorrFun} coincide up to a constant if and only if \eref{eq:xPert}  is justified.  So the failure of \eref{eq:xPert} leads to failure of { the} long-time Markovian behavior of the system correlation functions.

	For arbitrary spectral density, which leads to { the} long-time Markovian behavior in the second order of perturbation theory in $ \lambda $, let us consider  its sum with the spectral density \eref{eq:specDenNonDiff} for all possible $ \chi \in (0,1) $. For such a combined spectral density  the long-time Markovianity is violated in the second order of perturbation theory in $ \lambda $, but preserved in the zeroth order. So for any spectral density leading to long-time Markovianity in the second order we have found { an}  infinite number of spectral densities for which long-time Markovianity is violated in the second order and valid only in the first order. In this sense it is possible to say that for almost all spectral densities such that the system dynamics is long-time Markovian in the zeroth order of perturbation theory it is not long-time Markovian in the next order of perturbation theory.

\end{document}